\documentclass{article}
\usepackage[T1]{fontenc}
\usepackage[utf8]{inputenc}
\usepackage{ismir} 
\usepackage{amsmath,cite,url}
\usepackage{graphicx}
\usepackage{color}

\usepackage{hhline}

\usepackage{amssymb}
\usepackage{pifont}
\newcommand{\cmark}{\textcolor{blue}{\ding{51}}}%
\newcommand{\xmark}{\textcolor{red}{\ding{55}}}%
\usepackage{multirow}
\usepackage[symbol]{footmisc}
\usepackage{booktabs}
\usepackage[table,xcdraw]{xcolor}

\usepackage{tikz}
\usepackage{longtable}

\usepackage{subcaption}

\definecolor{lightgray}{RGB}{230,230,230}

\newcommand*\circledblue[1]{\tikz[baseline=(char.base)]{%
            \node[shape=circle,fill=blue!20,draw,inner sep=1pt] (char) {#1};}}

\usepackage{enumitem}

\usepackage{dsfont}

\title{GD-Retriever: Controllable Generative Text-Music Retrieval with Diffusion Models}

\multauthor
{Julien Guinot$^{*,1,2}$ \hspace{1cm} Elio Quinton$^2$ \hspace{1cm} György Fazekas$^1$} { 
$^1$ Centre for Digital Music, Queen Mary University of London, U.K.\\
$^2$ Music \& Audio Machine Learning Lab, Universal Music Group, London, U.K.\\
{{\tt \small j.guinot@qmul.ac.uk}}
}

\def\authorname{Julien Guinot, Elio Quinton, György Fazekas}

\usepackage[bookmarks=false,pdfauthor={\authorname},pdfsubject={\papersubject},hidelinks]{hyperref}

\hypersetup{
    colorlinks=true,
    linkcolor=blue,
    citecolor=blue,
    filecolor=magenta,      
    urlcolor=blue,
    pdftitle={Overleaf Example},
    pdfpagemode=FullScreen,
    }

\begin{document}

\maketitle

\vspace{-5pt}
\begin{abstract}

Multimodal contrastive models have achieved strong performance in text-audio retrieval and zero-shot settings, but improving joint embedding spaces remains an active research area. Less attention has been given to making these systems controllable and interactive for users. In text-music retrieval, the ambiguity of freeform language creates a many-to-many mapping, often resulting in inflexible or unsatisfying results.

We introduce Generative Diffusion Retriever (GDR), a novel framework that leverages diffusion models to generate queries in a retrieval-optimized latent space. This enables controllability through generative tools such as negative prompting and denoising diffusion implicit models (DDIM) inversion, opening a new direction in retrieval control. GDR improves retrieval performance over contrastive teacher models and supports retrieval in audio-only latent spaces using non-jointly trained encoders. Finally, we demonstrate that GDR enables effective post-hoc manipulation of retrieval behavior, enhancing interactive control for text-music retrieval tasks.

\end{abstract}
%


\begin{figure}
    \centering
    \includegraphics[width=.91\linewidth]{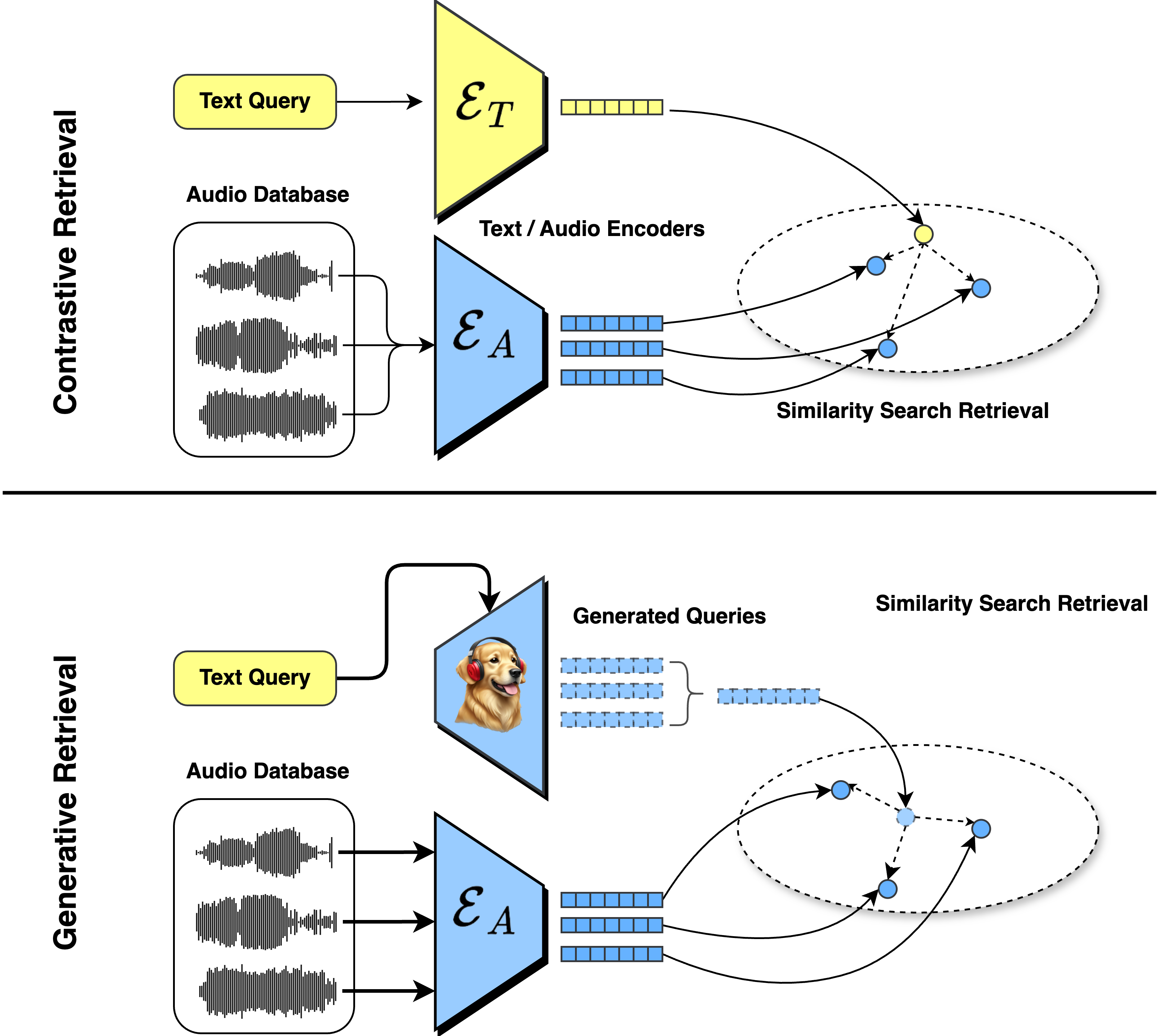}
    \caption{Overview of GD-Retriever's proposal: Instead of encoding text queries and audio keys in a joint embedding space (\textbf{top}), we generate queries in the audio space directly through conditioning on a text query (\textbf{bottom})}
    \label{fig:overview}
\end{figure}

\vspace{-5pt}

\section{Introduction}\label{Section: Introduction}

Multimodal text-music joint embedding models have largely facilitated text-queried music retrieval applications \cite{manco2022contrastive, mulan, clap, wu2024collap, elizalde2023clap}. Multimodal contrastive learning of text and music joint embedding spaces specifically have shown high performance on frozen probing tasks and promise for zero-shot classification approaches, with strong representation learning capacities.  Despite the improvement in effectively encoding musical information, these approaches often lack controllability. Unsatisfactory results from text querying require re-prompting with a different query to refine retrieval, and it is difficult to navigate the latent space of joint-embedding models with interpretable controls, such as ``\textit{I would like this retrieval result to be punchier}'' or ``\textit{I would like the retrieval result to be similar to this track, but with an electric guitar instead of acoustic}''.

One field in which such controls are more extensively explored, efficient, and disentangled is the field of generative AI. Diffusion generative models, specifically, have not only been widely adopted by virtue of their high quality outputs and multimodal conditionability \cite{ho2020denoising, audioldm, liu2023audioldm2, musicldm, musiccontrolnet, tango, nistal2024diff}, but have also been the focus of an extensive range of controllability approaches which have incrementally added powerful, multimodal, and intuitive controls to generative diffusion models \cite{musiccontrolnet, nistal2024diff, novack2024ditto, evans2024long, evans2024fast, gal2022image, mokady2023null, yang2023dynamic, baumann2024continuous, zhang2024compositional}. In light of this observation, we are strongly motivated to combine the retrieval and diffusion paradigms. This work focuses on exploring the capabilities of generative text-music models for retrieval, with the motivation of enabling controllability mechanisms for interactive retrieval. For instance, this would enable discovering directions of modification of musical attributes in the latent space or being able to modify the genre or instrumentation of a retrieved musical piece without modifying other semantic attributes.

We propose \textbf{G}enerative \textbf{D}iffusion \textbf{R}etriever, a new mechanism for retrieval, in which we train a conditional latent diffusion model on a retrieval-optimized latent space. We prompt GD-Retriever to generate hypothetical queries in the audio latent space and retrieve nearest neighbours (See Figure \ref{fig:overview}). Our contributions are\footnote{Code is made available at \href{https://github.com/Pliploop/GDRetriever}{https://github.com/Pliploop/GDRetriever}}:

\begin{enumerate}[label=\protect\circledblue{\arabic*},topsep=0pt,itemsep=-1ex,partopsep=1ex,parsep=1ex]
\item We present a generative diffusion retrieval framework that conditionally generates hypothetical queries in the latent space, leading to improved performance on in-domain text-music retrieval.
\item We distinguish ourselves from previous approaches by directly generating embedding sequences over aggregated embeddings, promoting finer-grained text-music understanding.
\item We show that we successfully unlock the array of controllability methods for generative models for retrieval through examples such as negative prompting and DDIM inversion \cite{mokady2023null}.
\item Our approach is directly applicable to audio-only latent spaces, and can leverage text encoders that have \emph{not} been jointly trained with an audio encoder.
\end{enumerate}
\vspace{-5pt}

\section{Background}\label{Section:Background}

\subsection{Text-music contrastive learning and retrieval}\label{Section:Background: CL}

Multimodal contrastive learning has shown strong results in computer vision \cite{cliptransferable, ho2020denoising, zhai2023sigmoid, bica2024improving}, and has been successfully extended to audio and music domains \cite{elizalde2023clap, manco2022contrastive, mulan}. These models encode paired text and audio inputs using encoders $\mathcal{E}_T$ and $\mathcal{E}_A$, project them into a shared latent space, and apply a contrastive InfoNCE loss \cite{chen2020simple} on pooled embeddings ($Z_T$, $Z_A$) to align positive pairs while separating negative ones.

Early text-audio/music models such as CLAP \cite{clap, elizalde2023clap}, MusCALL \cite{manco2022contrastive}, and MuLan \cite{mulan} adopt CLIP-style training \cite{cliptransferable}. Later work improves alignment through better captions and token-/time fine grained mechanisms \cite{wu2024collap, yuan2024t, manco2024augment, zhu2024cacophony}. Learned representations from these models are widely used for retrieval, in which the learned similarity metric between text and music encodings can be used to retrieve the most similar music key in a database of audio samples \cite{manco2022contrastive, mulan, clap}, generative conditioning \cite{audioldm, liu2023audioldm2, evans2024fast, evans2024long,comunitaSpecMaskGITMaskedGenerative2024}, and retrieval-augmented captioning \cite{li2025drcap, ghosh2024recap}.

\subsection{Diffusion Models} \label{Section: Background: Diffusion}

Diffusion models are powerful generative models that iteratively refine noisy inputs to generate high-quality outputs by learning a reverse Markov process \cite{ho2020denoising,esser2021taming,yu2022scaling}. These models corrupt data with noise over multiple steps and then learn to reconstruct the sample from the step information.

The denoising process is modeled as a learned transition, where at each step, a generator \( \mathcal{G} \) predicts either the original data \( x_0 \) (sample objective) or the noise added to the original latent ($\epsilon$ objective).
We adopt the sample prediction objective, where the model predicts the clean latent at each step, as in prior work \cite{ramesh2022hierarchical, mo2025diffgap}. 
The diffusion process can be conditioned on auxiliary conditioning information such as text or other modalities \cite{ho2022classifier,peebles2023scalable}, typically applied with classifier-free guidance (CFG) \cite{ho2020denoising} by interpolating between unconditional and conditional predictions at each denoising step. Latent diffusion models reduce computational costs by operating in a compressed latent space using pretrained autoencoders \cite{rombach2022high, audioldm, liu2023audioldm2, mousai, evans2024fast, evans2024long}.
\vspace{-5pt}

\subsection{Controllability for generation and retrieval}

Controllability in generation refers to how well generative models respond to human-guided interactions, allowing for attribute modification or refinement either during or after generation. In diffusion models, this includes techniques like text-based attribute editing \cite{dong2023prompt,hertz2022prompt,sridhar2024prompt}, inpainting \cite{lugmayr2022repaint}, inversion, and negative prompting \cite{miyake2023negative, mokady2023null}. In music generation, controllability is an active area of research due to its potential for creative applications \cite{zhang2024musicmagus,nistal2024diff,lin2024arrange}.

While extensively studied in generation, controllability in retrieval—especially for music—remains underexplored. This involves enabling users to guide or modify retrieval results interactively,  by specifying attributes of interest for retrieval in a disentangled way \cite{lee2020disentangled, guinot2025leave} or applying latent transformations, e.g. tempo adjustments \cite{mccallum2024similar}. Generative retrieval has emerged in recent work to generate latent queries, mainly to improve performance in general audio retrieval \cite{mo2025diffgap} or add multimodal guidance \cite{bao2025diff4steer}, rather than enabling interactivity during or after retrieval.

\vspace{-2pt}

\begin{figure*}[t]
    \centering
    \hspace{-20pt}
    \begin{subfigure}{0.5\textwidth}
        \centering
        \includegraphics[height=4.6cm]{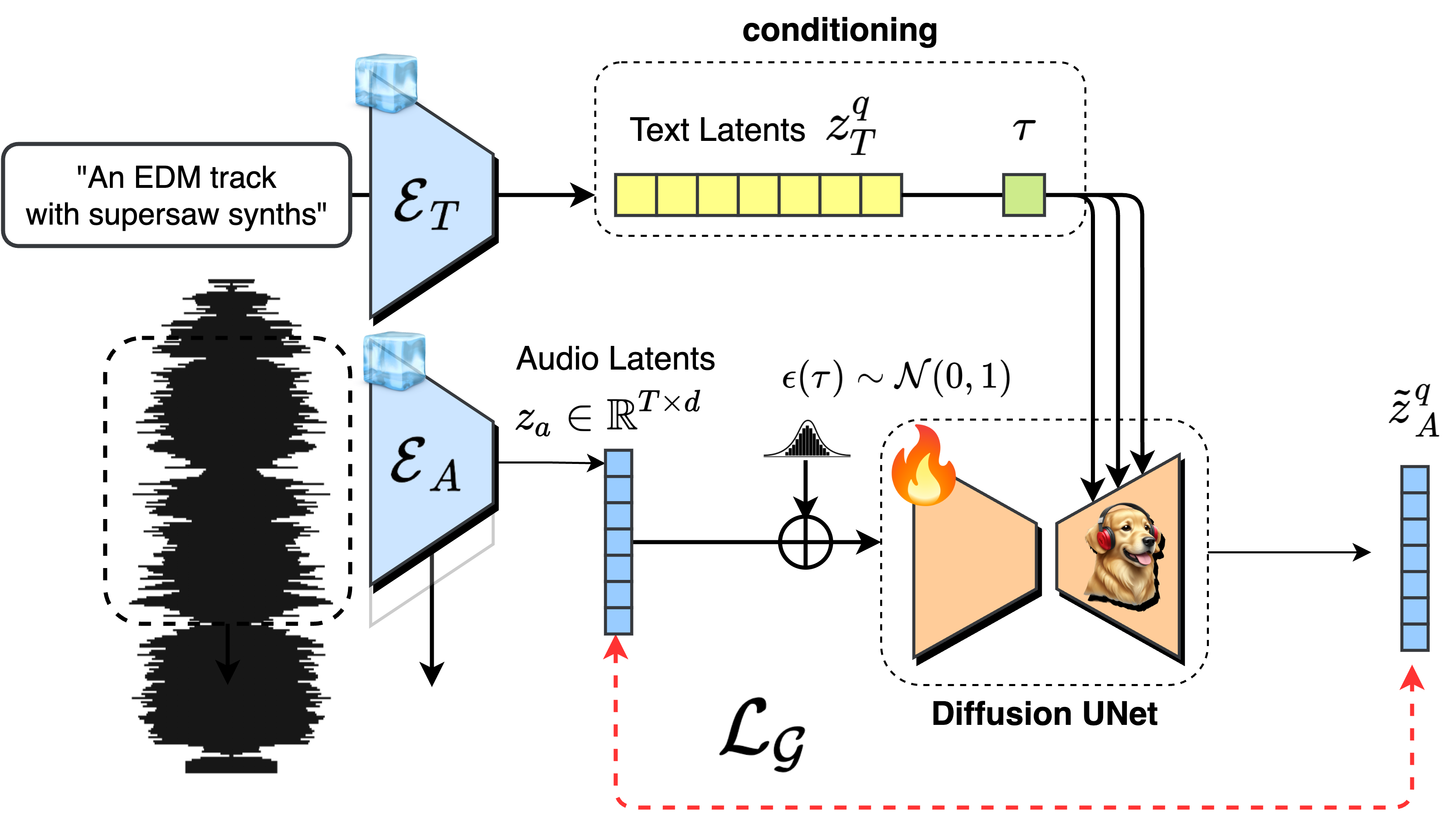} 
        \caption{\textbf{Stage 1 --- Generative pretraining}. }
        \label{fig:sub1}
    \end{subfigure}
    \hspace{0.02\textwidth} 
    \vrule width 0.5pt 
    \hspace{0.02\textwidth} 
    \begin{subfigure}{0.4\textwidth}
        \centering
        \includegraphics[height=4.6cm]{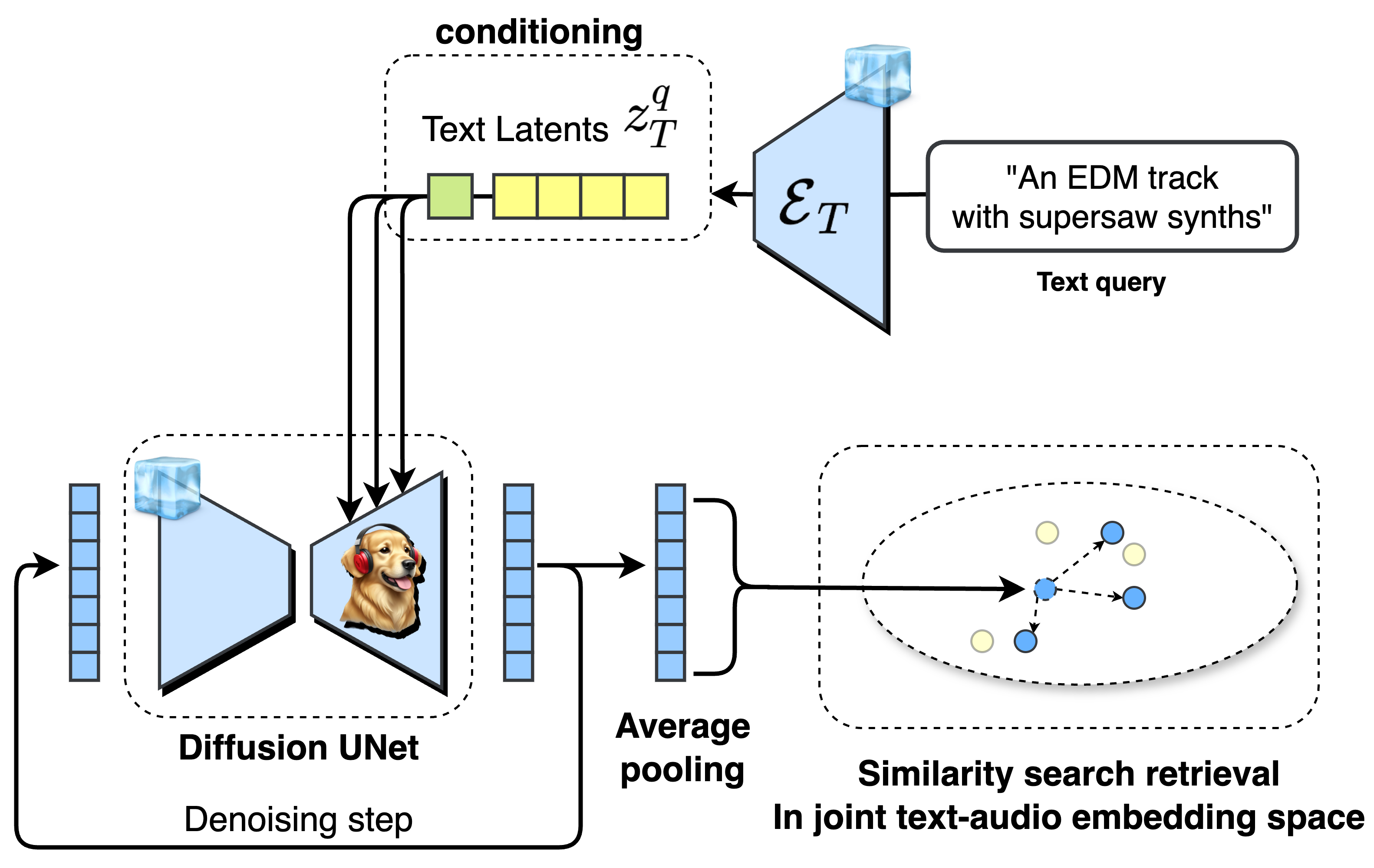} 
        \caption{\textbf{Stage 2 --- Retrieval}}
        \label{fig:sub2}
    \end{subfigure}
    \caption{\textbf{GD Retriever Method}: We train a model to generate text-conditioned ghost queries for retrieval. \textbf{Left}: A diffusion model is trained to generate audio latents from text captions. \textbf{Right}: Using the frozen model, we generate audio embeddings from a caption to retrieve similar audio via ghost queries.}
    \label{fig:approach}
\end{figure*}

\section{Generative Diffusion Retrieval}\label{Section:Methods}


We propose an intuitive generative approach to retrieval using diffusion models, which we name \textbf{G}enerative \textbf{D}iffusion \textbf{R}etriever. Using a pretrained latent space optimized for audio-audio retrieval, we train a generative diffusion model conditioned on text to generate audio latent embeddings in this space. At inference time, rather than encoding the text query into the shared latent space and retrieving the nearest audio \cite{manco2022contrastive, clap}, we generate a ``ghost'' audio query in the latent space conditioned on the text query and retrieve the nearest neighbours in the audio space.
By using a generative model as a retriever, we enable adaptation of audio-only latent spaces for text-audio retrieval without requiring multimodal pretraining. The generative modelling of retrieval allows for greater retrieval controllability through attribute modification and interactive refinement techniques from the generative domain. The approach is shown in Figure \ref{fig:approach}.


Consider an audio, query caption pair $\{x_t^q, x_a\}$ and a conditioning text encoder $\mathcal{E}_T$ which encodes the text query into a sequence of embeddings $z_T^q$. Let $\mathcal{E}_A$ be a pretrained and frozen audio encoder encoding $x_a$ into a sequence of audio embeddings $z_A$.  We train $\mathcal{G}$ with a diffusion loss to reconstruct $z_A$ conditioned on $z_T^q$, which we notate $\Tilde{z}_A^q$:


\begin{equation}
    \mathcal{L}_\mathcal{G} = \mathbb{E}_{\tau, Z_a,z_T^q} \left[ \left\| z_a - \mathcal{G}(z_{a,\tau}, \tau,z_T^q) \right\|_2^2 \right]
\end{equation}

Where $\tau$ is the diffusion step. We aggregate $\tilde{z}_A^q$ into $\tilde{Z}_A^q$ through averaging over the sequence timesteps and use $\tilde{Z}_A^q$ as a query in the audio space to retrieve audio.

\vspace{-3pt}




\subsection{Model architecture}\label{Section: Model architecture}

\vspace{-1pt}

\subsubsection{Diffusion backbone}\label{Section:diffusion backbone}

We use a well-established UNet with cross-attention conditioning \cite{ho2020denoising, audioldm,liu2023audioldm2, musicldm} as our diffusion model. Early experimentation led to the design choice of a $\sim$40M parameter model. Models are conditioned on text embedding sequences through cross-attention. Compared to previous work \cite{mo2025diffgap,bao2025diff4steer} that performs generative diffusion retrieval with aggregated embeddings, this enables more fine-grained interaction between audio and text \cite{bao2025diff4steer}, as we show in Section \ref{subsection:quality of generated queries}. Supported by results in \cite{ramesh2022hierarchical}, we find that a sample-objective (See Section \ref{Section:Background}) yields better retrieval results than $\epsilon$-objective. \vspace{-5pt}

\subsubsection{Text and audio encoders}\label{Section: Encoders}


We use three audio encoders: CLAP \cite{clap}, MusCALL \cite{manco2022contrastive}, and MULE \cite{mccallum2022supervised}. CLAP uses an HTSAT backbone \cite{chen2022hts} with the publicly available \textit{Music} checkpoint\footnote{\hyperlink{https://github.com/LAION-AI/CLAP}{https://github.com/LAION-AI/CLAP}}. MusCALL is based on a ResNet50 encoder, and MULE is reimplemented and pretrained on MTG-Jamendo following \cite{guinot2025leave}. For text encoders, we use CLAP’s RoBERTa-based encoder \cite{chung2024scaling, liu2019roberta}, MusCALL’s 4-layer transformer, or a pretrained Flan-T5 model \cite{chung2024scaling} from HuggingFace\footnote{\hyperlink{https://huggingface.co/google/flan-t5-base}{https://huggingface.co/google/flan-t5-base}}. Flan-T5 is a fine-tuned T5 language model commonly used in music generation \cite{tango, melechovsky2023mustango, liu2023audioldm2, copet2023simple}.

\vspace{-4pt}

\subsection{Datasets}\label{subsection:dataset}

We use two well-explored public music-caption pairs datasets as well as a private dataset for training and evaluation. Song Describer \cite{manco2023song} (SD) is a dataset of 1100 crowd-sourced captions corresponding to 700 excerpts of 2 minute music clips. MusicCaps \cite{agostinelli2023musiclm} is another music-caption pair dataset consisting of 5500 pairs with 10s audio. Finally, we use a private large scale dataset of professionally-annotated song descriptions (PrivateCaps). Table \ref{tab:datasets} inventories dataset scales. For evaluation on PrivateCaps, we use a 5500-sample subset of the test set.

\begin{table}[h]
\centering
\resizebox{1\columnwidth}{!}{%
\begin{tabular}{llllll} \hline
\textbf{Dataset} & \textbf{\#tracks} & \textbf{\#captions} & \textbf{Hours} & \textbf{Training} & \textbf{Eval} \\ \hline
SongDescriber \cite{manco2023song} & 0.7k & 1.1k & 23.3 & \xmark & \cmark \\
MusicCaps \cite{copet2023simple} & 5.5k & 5.5k & 15.3 & \xmark & \cmark \\
PrivateCaps & 251k & 251k & 12.5k & \cmark & \cmark \\ \hline
\end{tabular} 
}
\caption{Dataset details - PrivateCaps is an internal dataset of full-length professionally annotated production tracks}
\label{tab:datasets}
\end{table}

\vspace{-10pt}

\subsection{Training details}\label{Section: Details}

We extract latent embeddings from our training datasets (See Section \ref{subsection:dataset}) with the selected audio encoders. We extract $z_a$ by sliding the encoder over the audio input at a frequency of 1Hz. GDR is trained on PrivateCaps with a batch size of 256 for 100k steps, with an early stopping mechanism on validation diffusion loss. Models are trained on a single A5000 GPU. We use AdamW and
linear warmup the learning rate to $1\mathrm{e}{-4}$ over 5000 steps then cosine decay to 0. Our model is trained on latent sequences of length $T=64$, i.e. 1 minute of audio. 
We use CFG on text conditioning \cite{ho2022classifier} (masking probability 10\%).

\vspace{-5pt}



\section{Experiments}\label{Section: Experiments}

\subsection{Retrieval}

\begin{table}[h!]
\centering
\centering
\resizebox{.79\columnwidth}{!}{%
\begin{tabular}{llllccc}
\hline
\multicolumn{2}{l}{T$\rightarrow$A } &  & \multicolumn{4}{c}{Eval dataset} \\ \cline{1-2} \cline{4-4} \cline{6-7} 
Model & Metric &  & PC &  & SD & MC\\ \hline
\multirow{4}{*}{CLAP} & R@1 $\uparrow$ &  & 2.2 &  & 3.1 & \textbf{3.8} \\
 & R@5 $\uparrow$ &  & 7.2 &  & 13.7 & \textbf{12.9} \\
 & R@10 $\uparrow$ &  & 12.3 &  & 23.2 & \textbf{19.5} \\
 & MedR (\%) $\downarrow$ &  & 3.7 &  & 4.0 & \textbf{1.4} \\ \hline
 \multirow{4}{*}{GDR-CLAP} & R@1 $\uparrow$ &  & \textbf{6.9} &  & \textbf{4.7} & 2.7 \\
 & R@5 $\uparrow$ &  & \textbf{17.1} &  & \textbf{15.3} & 7.6 \\
 & R@10 $\uparrow$ &  & \textbf{22.9} &  & \textbf{24.7} & 11.5 \\
 & MedR (\%) $\downarrow$ &  & \textbf{1.6} &  & \textbf{3.8} & 2.9 \\ \hline 
\addlinespace[-8pt]
 \\ \hline

 \multirow{4}{*}{MusCALL} & R@1 $\uparrow$ &  & 10.1 &  & 3.6 & 1.0 \\ 
 & R@5 $\uparrow$ &  & \textbf{26.2} &  & 13.6 & 3.9 \\
 & R@10 $\uparrow$ &  & \textbf{35.1} &  & 22.0 & 7.0 \\
 & MedR (\%) $\downarrow$ &  & \textbf{0.4} &  & 4.2 & 5.1 \\ \hline
 \multirow{4}{*}{GDR-MusCALL} & R@1 $\uparrow$ &  & \textbf{10.8} &  & \textbf{5.1} & \textbf{1.8} \\
 & R@5 $\uparrow$ &  & 25.1 &  & \textbf{16.9} & \textbf{6.4} \\
 & R@10 $\uparrow$ &  & 33.3 &  & \textbf{25.5} & \textbf{9.9} \\
 & MedR (\%) $\downarrow$ &  & 0.6 &  & \textbf{3.5} & \textbf{3.4} \\ \bottomrule 
\end{tabular}%
}
\caption{Main retrieval results for GDR. We compare GDR-CLAP to CLAP and GDR-MusCALL to muscall for R@1,5,10 on the PC, SD and MC Datasets.}
\label{tab:main}
\vspace{-5pt}
\end{table}

We evaluate GD-Retriever’s retrieval performance against teacher models. We generate $n_q = 5$ audio latent queries $\tilde{z}_A$ conditioned on $z_T^q$ We average $\tilde{z}_A$ over time and $n_q$ into $\Tilde{Z}_A$ for retrieval. Teachers encode text and audio into embeddings $Z_T$ and $Z_A$. . Results are shown in Table \ref{tab:main}.
While GD-Retriever outperforms teacher models in several in-domain scenarios—most notably on PrivateCaps (PC) and to a lesser extent SongDescriber (SD)—its retrieval performance degrades on out-of-domain MusicCaps (MC). In particular, GDR-CLAP underperforms on MC relative to the CLAP teacher, despite showing strong improvements on PC. Conversely, GDR-MusCALL underperforms on PC while providing stronger performance than the teacher baseline on MC and SD. Despite strong in-domain results, these inconsistencies suggest that domain mismatch plays a role in limiting retrieval performance.

\subsubsection{Domain adaptation}
\label{sec:domain-adaptation}



We identify two compounding sources of domain shift. First, pretrained contrastive models like CLAP and MusCALL often fail to generalize across datasets with differing audio and text distributions—a well-known issue in dense retrieval and multimodal learning \cite{yu2022cocodr, khramtsova2023selecting}. CLAP, trained on LAION-630k \cite{clap}, performs worse than MusCALL on its in-domain evaluation set (PC), but shows similar performance on SD, and better performance on MC.

Second, GD-Retriever learns a text-to-audio mapping on the teacher’s frozen embedding space during diffusion training. If the teacher suffers from domain mismatch (e.g., CLAP on PC), GDR can compensate by adapting to the training distribution. Conversely, when the teacher is well-aligned (e.g., MusCALL on PC), GDR tends to match, but not exceed, its performance. This explains why GDR-CLAP, trained on PrivateCaps with a CLAP teacher, reproduces performance trends seen in MusCALL’s teacher embeddings—performing best on PC, acceptably on SD, and poorly on MC.

\begin{table}[h]
\centering
\resizebox{\columnwidth}{!}{
\begin{tabular}{l l | ccc | cc  cc}
\toprule
Dataset & Encoder Pair 
        & FTD ↓ & FAD & R@5 
        & \multicolumn{2}{c}{FAD ↓} 
        & \multicolumn{2}{c}{R@5 ↑} \\ 
        & 
        & $Z_T$ 
        & $Z_A$ & $Z_A$ & $\tilde{Z}_A$
        & $\tilde{Z}_A^{\text{align}}$  & $\tilde{Z}_A$ & $\tilde{Z}_A^{\text{align}}$ \\
\midrule
PC & CLAP     &   -    &   -    &   7.2    &   0.03    &  0.003     &  17.1     &  18.2   \\
SD & CLAP     &   -    &    -   &    13.7   &  0.09    &  0.001      &   15.3    &   15.9    \\
MC & CLAP     &   -    &   -    &   12.9    & 0.34     & 0.001 &      7.6 & 8.1      \\
\midrule
PC & MusCALL  &     0.008  &   0.002    &  26.2  &     0.02  &   $\sim$0    &    25.1   &   25.3    \\
SD & MusCALL  &   0.14   &   0.18    & 13.6      &  0.12     &     $\sim$0   &    16.9   &   17.7    \\
MC & MusCALL  &   0.32    &  0.20     &  3.9     &  0.17    &    $\sim$0   &     6.4  & 7.2      \\
\bottomrule
\end{tabular}
}
\caption{Fréchet distances of $Z_A/Z_T$ to the training distribution of teacher models, Fréchet audio distance (FAD) to the evaluation set, and retrieval performance (R@5) for generated queries ($\tilde{Z}_A$), and aligned queries ($\tilde{Z}_A^{\text{align}}$) across datasets and encoder pairs. We are unable to evaluate FAD on $Z_A$ for CLAP's training set as LAION-630k is private.}
\label{tab:alignment}
\end{table}

To test this hypothesis, we compute Frechet Distances between training and evaluation distributions, comparing (1) teacher audio embeddings of each evaluation set to the training distribution and (2) GDR-generated queries to the evaluation set, before and after a lightweight alignment step. Following prior domain adaptation work \cite{zhou2023test, sun2017correlation}, we apply a post-hoc shift in mean and covariance to $\tilde{Z}_A$ to match the evaluation set (notated  $Z_A^{\text{align}}$). This method is model-agnostic, efficient, and requires no retraining.

We evaluate on joint encoder pairs to illustrate the joint text-audio domain generalization gap. As shown in Table~\ref{tab:alignment}, GDR-generated latents exhibit similar distribution shifts as the teacher. Alignment consistently reduces FAD and improves R@5, supporting our claim that retrieval degradation stems from inherited distribution divergence rather than a limitation of our approach. While not a complete solution, this provides both evidence for our diagnosis and a simple, effective mitigation. We leave broader generalization strategies for future work. 
\vspace{-3pt}

\subsubsection{Encoder pair variation}\label{Section: encoder pair variation}


Two core affordances of GDR are its ability to (1) operate in audio-only latent spaces not trained jointly with text, and (2) support arbitrary text encoders for conditioning. This is enabled by the diffusion model learning a generative mapping between text and audio embeddings, independent of any contrastive pre-training alignment. The generative retrieval objective imposes no constraints on the multimodality of the space or the choice of text encoder. To demonstrate this, we test several combinations of audio and text encoders that were not jointly trained: we replace the text encoder in GDR-CLAP with Flan-T5 (Section~\ref{Section: Encoders}), and use the audio encoder to MULE paired with T5. Retrieval results are reported in Table~\ref{tab:combined}.

\begin{table}[h]
\centering
\resizebox{.79\columnwidth}{!}{
\begin{tabular}{lllcccc}
\hline
\multicolumn{3}{c}{T$\longrightarrow$A} &  & \multicolumn{3}{c}{Eval dataset} \\ \cline{1-4} \hline
Model  & $\mathcal{E}_T$ & Metric &  & PC & SD & MC \\ \hline
\multirow{7}{*}{GDR-CLAP} 
  & \multirow{4}{*}{T5} & R@1 $\uparrow$ &  & \textbf{8.1} & \textbf{4.9} & 2.3 \\
  &                     & R@5 $\uparrow$ &  & \textbf{21.1} & \textbf{15.6} & \textbf{7.8} \\
  &                     & R@10 $\uparrow$ & & \textbf{29.2} & \textbf{25.1} & \textbf{11.7} \\
  &                     & MR $\downarrow$ &  & \textbf{0.8} & \textbf{3.7} & 2.9 \\ \cline{2-7}
  & \multirow{3}{*}{CLAP} & R@1 $\uparrow$ &  & 6.9 & 4.7 & \textbf{2.7} \\
  &                     & R@5 $\uparrow$ &  & 17.1 & 15.3 & 7.6 \\
  &                     & R@10 $\uparrow$ & & 22.9 & 24.7 & 11.5 \\
  &                     & MR $\downarrow$ & & 1.6 & 3.8 & 2.9 \\ \hline
\multirow{4}{*}{GDR-MULE}  
  & \multirow{4}{*}{T5} & R@1 $\uparrow$ &  & 7.6 & 4.1 & 1.6 \\
  &                     & R@5 $\uparrow$ &  & 18.5 & 13.9 & 6.2 \\
  &                     & R@10 $\uparrow$ & & 25.3 & 21.8 & 11.0 \\
  &                     & MR $\downarrow$ & & 1.6 & 4.2 & 3.2 \\ \hline
\end{tabular}%
}
\caption{Comparison of retrieval performance across models and text encoders. GD-Retriever enables text-conditioned retrieval on non-jointly trained encoders.}
\label{tab:combined}
\end{table}



Including T5 as the text encoder improves in-domain performance for GDR-CLAP, likely by regularizing the mapping between contrastively trained audio and text encoders. However, this benefit is more limited out-of-domain, consistent with our findings in Section~\ref{sec:domain-adaptation}, where domain mismatch stems from training set distributions. We also find that GDR-MULE supports text-music retrieval and outperforms the CLAP teacher on PrivateCaps and SongDescriber. This demonstrates that joint retrieval spaces can be built from unimodal audio latents without large-scale multimodal pre-training—a key affordance of our approach.

\vspace{-3pt}

\subsection{Quality of Generated Queries}\label{subsection:quality of generated queries}

Beyond retrieval, we evaluate generated query quality using fidelity, diversity, and prompt adherence metrics. FAD captures audio fidelity, while CLAP score assesses alignment with input text. To evaluate diversity, we generate clusters of 10 audio queries per prompt and measure the intrasample cosine similarity (MICS) following previous work \cite{bao2025diff4steer} and the normalized Vendi score (NVendi), along with its intracluster variant (MINVS), to assess the variation in generated queries \cite{friedman2022vendi}.

We compare our diffusion-based UNet to two baselines: a regression UNet predicting sequences of audio embeddings from sequences of learned mask embeddings conditioned on $z_T^q$, and two 1-hidden-layer MLP (GeLU, 4096 units), trained with respectively diffusion and regression objectives to reconstruct $Z_A$ conditioned on $Z_T^q$. All models use the same training hyperparameters and CLAP T/A encoders. Table~\ref{tab:fidelityetc} shows results on PrivateCaps. Our diffusion UNet outperforms alternatives in retrieval and fidelity. While MLP Diffusion achieves higher diversity, it comes at the cost of realism and retrieval performance.

\begin{table}[h]
\centering
\resizebox{.88\columnwidth}{!}{%
\begin{tabular}{llcc|cc}
\hline
\multirow{2}{*}{\textbf{}} & \multirow{2}{*}{\textbf{Metric}} & \multicolumn{2}{c|}{\textbf{Diffusion}} & \multicolumn{2}{c}{\textbf{Regression}} \\
 & & \textbf{UNet} & \textbf{MLP} & \textbf{UNet} & \textbf{MLP} \\
\hline
\multirow{2}{*}{Retrieval (SD)} & R@5 $\uparrow$ & \textbf{15.3} & 11.3 & 9.1 & 7.1 \\
 & NMedR $\downarrow$ & \textbf{3.8} & 5.5 & 5.6 & 5.1 \\
\hline
\multirow{2}{*}{Fidelity} & FAD $\downarrow$ & \textbf{0.09} & 0.13 & 0.13 & 0.13 \\
 & CLAP $\uparrow$ & \textbf{0.63} & 0.53 & 0.54 & 0.53 \\
\hline
\multirow{3}{*}{Diversity} & MICS $\downarrow$ & 0.92 & \textbf{0.62} & 1 & 1 \\
 & MINVS $\uparrow$ & 1.42 & \textbf{4.53} & 1 & 1 \\
 & NVendi $\uparrow$ & 12.9 & \textbf{68.9} & 3 & 3.4 \\
\hline
\end{tabular}%
}
\caption{Fidelity, Retrieval, and Diversity metrics for CLAP-Retrievers trained on PrivateCaps. FAD is grounded on MTG-Jamendo \cite{bogdanov2019mtg}.}
\label{tab:fidelityetc}
\end{table}

\vspace{-15pt}

\subsection{Controllability}

\subsubsection{Negative prompting}

Negative prompting is an off-the-shelf affordance of diffusion models related to CFG \cite{ho2022classifier}, which allows users to specify what they do \emph{not} want to be in the generated output. Negative prompting modifies the CFG update by incorporating a negative conditioning signal \( z_T^{q-} \). Given a query embedding \( z_T^{q+} \), a denoising step is given by:

\vspace{-2pt}

\begin{equation}
\begin{split}
    \tilde{z}_{A,\tau+1}^{NP} = (1+w)\mathcal{G}(\tilde{z}_{A,\tau}, \tau+1, z_T^{q+}) \\ - w\mathcal{G}(\tilde{z}_{A,\tau},\tau+1, z_T^{q-})
    \end{split}
\label{negative}
\end{equation}
where $w$ is the classifier free guidance strength \cite{ho2022classifier}.  
This formulation removes undesired attributes by interpolating towards conditional generation and away from negatively conditioned outputs at each diffusion step (See Section \ref{Section:Methods}).

We evaluate the effectiveness of negative prompting in retrieval by curating 50 negative prompts across \textit{genre, mood, instrumentation, key, and tempo} (e.g. ``\textit{a rock song}'' as a negative prompt ``removes'' rock). Each category includes different phrasing formats.
For each query $q$, we create new, modified query latents $z_a^{mod}$ using negative prompting and three additional modification methods as baselines, all applied with guidance strength $w$: 

Negative prompting (NP) modifies latents by applying Eq. \ref{negative} with $z_T^-$ as negative conditioning. Text Interpolation (\(\Delta_T\)) interpolates from $Z_A$ away from $Z^{T^{q-}}$: \( Z_A' = Z_A + w\Delta_T \), where $\Delta_T = Z_T^{q+} - Z_T^{q-}$. Audio interpolation interpolates along the direction from $\tilde{Z}_A$ and away from $\tilde{Z}_A^-$ : \( \Delta_A = \tilde{Z}_A - \tilde{Z}_A^- \). Our last baseline is Prefix Negation Prompting (PNP): We modify the query by negating attributes (e.g., ``a rock track'' → ``\textit{not} a rock track''), then generate $\tilde{z}_A^{PNP}$ with GDR.

\begin{figure}[]
    \centering
        \includegraphics[width=\columnwidth]{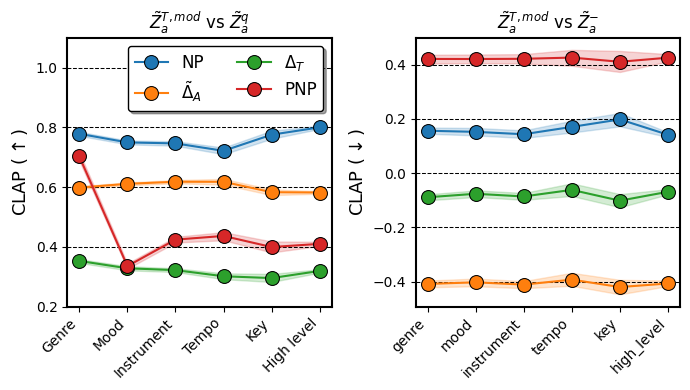} 
        \label{fig:positive_negprompt}
    \vspace{-15pt}
    \caption{CLAP score results between $Z_a^{mod}$ and $\tilde{Z}_A^{q+}$, $Z_a^{mod}$ and $\tilde{Z}_a^{q-}$ for negative prompting and baselines for different categories.}
    \label{fig:negprompt}
    \vspace{-5pt}
\end{figure}

$Z_a^{mod}$ obtained from all modification methods is compared using CLAP score to $Z_A^{q+}$, $\tilde{Z}_A^{q+}$, and $Z_A^{q-}$, for which a higher CLAP score is better as it signifies higher similarity to the positive prompt. We also compare $Z_a^{mod}$ to $\tilde{Z}_A^{q-}$, and $Z_A^{q-}$, for which a lower CLAP score (less similar to the negative conditioning) is better. We also use FAD to assess the fidelity of $Z_A^{mod}$. A modified audio latent that is similar to the original $Z_A^{q+}$ and dissimilar to $Z_A^{q-}$ but is very far from any reasonable distribution (i.e. large FAD) will yield unrealistic or uncommon music results.
CLAP score and FAD grounded on MTG-Jamendo are reported Table \ref{tab:negative_prompting}. Fine-grained category experiments are shown Table \ref{fig:negprompt}.

\begin{table}[h]
\centering
\resizebox{.85\columnwidth}{!}{%
\begin{tabular}{lllccccccc} \toprule
\multicolumn{2}{c}{key} &  & & & $Z_A^{q+}$  & \multicolumn{4}{c}{Modified query $\tilde{Z}_A^{mod}$} \\ \cline{1-2} \cline{5-10} 
\multicolumn{2}{l}{} &
   &
   &
   &
   &
  $\text{NP}$ &
  $\Delta_T$ &
  $\tilde{\Delta}_A$ &
  $\text{PNP}$ \\\cline{1-10}
\multirow{3}{*}{$\text{CLAP} \uparrow$}   & $Z_A$           &  &     &  &   0.69    &  \textbf{0.66}&  0.38    &0.39  &0.51  \\
                              & $\tilde{Z}_A$   &  &     & &    1       &  \textbf{0.85}&   0.65   &0.62  &0.49  \\
                              & $Z_T$           &  &     &  &   0.42    &  \textbf{0.39}&   0.28   &0.33  &0.21  \\ \hline
  \multirow{2}{*}{$ \text{CLAP} \downarrow$} & $\tilde{Z}_A^n$ &  &     &  &   0.41    &  0.21&   -0.02   & \textbf{-0.23}  &0.46  \\
                              & $Z_T^n$         &  &     &  &   0.17    &  0.08&   \textbf{-0.51}   & -0.04  &0.21  \\\hline
$\text{Fidelity}$               & FAD             &  &    &   & 0.11      & \textbf{0.12}  &  3.12    & 0.60  &  \textbf{0.12} \\\bottomrule
\end{tabular}%
}
\caption{Negative prompting experiments - CLAP score of modified queries $z_A^{mod}$ vs original/negative $Z_A/Z_T$}
\label{tab:negative_prompting}
\end{table}

Negative prompting shows strong desirable results: The modified audio latent remains the most similar to the original prompt across modification baselines, while significantly distancing itself from the negative prompt. At the same time, NP-modified latents remain realistic and in distribution, as demonstrated by the lower FAD score. While $\Delta_T$ and $\Delta_A$ can lower the CLAP score relative to undesired attributes, they also degrade semantic alignment with the original prompt, as reflected in lower CLAP scores. In addition, this reduction is achieved unrealistically: higher FAD scores indicate the resulting latents deviate significantly from the grounding distribution. In retrieval, this would lead to unnatural or implausible results which are less relevant, even if less similar to the negative attribute. Moreover, a lower CLAP score to a negative prompt is not always desirable: Usually, a CLAP score of 0 denotes the absence of shared information between the two embeddings, while a negative CLAP score signifies opposite information. Colloquially, a CLAP score of -1 for a reference embedding of \textit{''guitar``} does not mean {''\textit{No guitar}``}, but rather \textit{''the opposite of guitar''}, which is ill-defined.



\subsubsection{DDIM Inversion}


One useful property of diffusion models is DDIM inversion, which enables user-controlled modifications while preserving semantic similarity \cite{mokady2023null}. By re-noising an embedding via an inverse DDIM scheduler, the latent returns to a noisier state where high-level features are established. From this pivot, applying new guidance during denoising yields outputs that better match the new prompt while remaining close to the original.

This is valuable for retrieval, where users may want to refine a specific attribute of a query they are partly satisfied with. Current joint embedding retrieval models lack native support for such interaction. In contrast, GD-Retriever supports DDIM inversion directly, allowing for interactive and controllable retrieval refinement.

\begin{figure}[h]
    \centering
    \includegraphics[width=1\columnwidth]{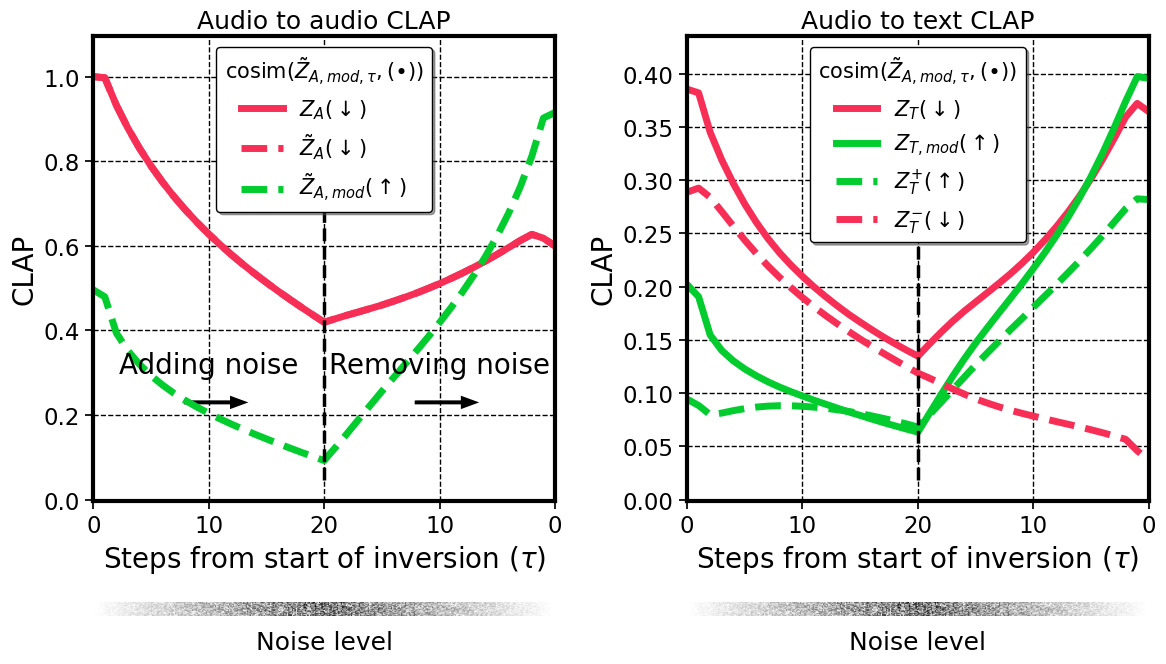}
    \caption{DDIM inversion example. \textbf{Left}: CLAP scores to audio from original and modified prompts. \textbf{Right}: CLAP scores to text prompts and added/removed words.}
    \label{fig:inversionexample}
\end{figure}

To demonstrate DDIM inversion for retrieval, we apply it to a prompt from the Song Describer dataset: ``A \textcolor{red}{choppy beat-heavy} acoustic guitar song with \textcolor{red}{soft} vocals.'' From the generated latent $\tilde{z}_A^q$, we perform inversion using an inversion prompt $x_T^{q, \text{inv}}$: ``A \textcolor{blue}{smooth, solo} acoustic guitar song with \textcolor{blue}{harsh} vocals.'' The aim is to produce a latent $Z_A^{mod}$ that remains close to the original while aligning more with the modified prompt. We track CLAP scores between $Z_A^{mod}(\tau)$ with $\tau$ the inversion step and both original and modified audio/text queries: $\tilde{Z}_A^q$, $\tilde{Z}_A^{q,\text{inv}}$, $Z_T^q$, $Z_T^{q\text{inv}}$, and the text encodings of added ($Z_T^{+}$) and removed ($Z_T^{-}$) words. Results are shown in Figure~\ref{fig:inversionexample}. We observe a clear transition: CLAP similarity shifts from the original to the modified prompt while remaining high to the original, confirming the effectiveness of DDIM inversion for fine-grained control in retrieval.

To validate the usability of DDIM inversion as a retrieval controllability tool on a larger scale. We now curate 50 prompts from the Song Describer dataset that would represent realistic use-cases for refining a search result for retrieval, and curate modified prompts representing realistic modifications to refine a query, either by modifying qualificatives or subjects, or adding more details.

\begin{figure}[h]
    \centering
    \includegraphics[width=\columnwidth]{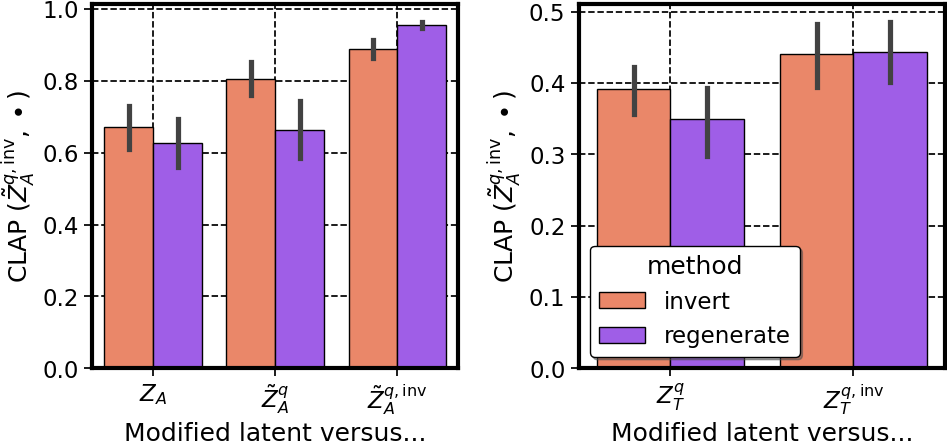}
    \caption{Systematic evaluation of DDIM inversion on curated prompt modifications, comparing CLAP score of inverted latents and regenerated latents to (\textbf{left}: audio, \textbf{right}: text) original and modified latents.}
    \label{fig:DDIMlarge}
\end{figure}

Original prompts are notated $z_T^q$ and modified prompts $z_T^{q'}$ (again, modified audio latents are notated $z_a^{mod}$). For inversion, we re-noise $\tilde{z}_A^{q}$ for 20 out of 50 steps and denoise \ conditioned on $z_T^{q'}$. We use re-generation as a baseline by simply generating $\tilde{z}_A^{q'}$. We compare $Z_a^{mod}$ to $Z_A^q$, $\tilde{Z}_A^q$ and $\tilde{Z}_A^{q'}$ for audio comparison, and $Z_T^q$, $Z_T^{q'}$ for text. A desirable result is for $Z_A^{mod}$ to be similar to these embeddings, meaning semantic similarity to  the original and modified prompts. Results are shown in Figure \ref{fig:DDIMlarge}.

\vspace{-5pt}


DDIM inversion, a native affordance of GDR, yields higher similarity to the original prompt compared to regeneration, which causes a drop in CLAP score between $Z_A^{\text{regen}}$ and $\tilde{Z}_A^{q'}$. In contrast, $Z_A^{\text{inv}}$ remains close to $\tilde{Z}_A^{q}$. While both methods maintain similar alignment with the modified prompt, only inversion preserves semantic similarity to the original, making it a more faithful controllability mechanism. Additionally, we observe that both classifier-free guidance and the number of inversion steps modulate the strength of the effect, offering further control. While we use vanilla DDIM here, recent improvements in inversion techniques \cite{mokady2023null, dong2023prompt} can be directly applied to GDR for more realistic and fine-grained control.










\vspace{-2pt}

\section{Conclusion and future work}\label{Section: conclusion}



We present GD-Retriever, a generative framework for text-to-music retrieval that uses diffusion models to produce latent queries in retrieval-relevant spaces. GD-Retriever outperforms contrastive teacher models on in-domain data and enables retrieval in unimodal audio spaces by leveraging independently pretrained text and audio encoders—removing the need for joint multimodal training.

Beyond retrieval performance, GD-Retriever enables inference-time controllability through negative prompting and DDIM inversion, offering flexible and post-hoc manipulation of retrieval behavior. These affordances open the door to interactive and user-steerable music retrieval. While domain mismatch remains a challenge, our findings suggest this can be partially mitigated through latent alignment. We encourage future work to expand generative control in retrieval, aiming for more robust, adaptable, and expressive retrieval systems.

\clearpage

\section{Acknowledgement}

This work is supported by the EPSRC UKRI Centre for
Doctoral Training in Artificial Intelligence and Music
(EP/S022694/1) and Universal Music Group.

\bibliography{ISMIRtemplate}

%
%
%
%

\clearpage

\end{document}